\documentclass{icsm}         
\usepackage{graphicx}
\usepackage{bm}
\usepackage{amsmath}
\usepackage{amsfonts}
\usepackage{amssymb}
\usepackage{amsthm}

\usepackage[dvips]{epsfig}

\begin{document}
\title{Combinatorial Solutions to Normal Ordering of Bosons}
\authori{P. Blasiak$^{a,b}$, A. Gawron$^{a}$, A. Horzela$^{a}$,}
\addressi{$^{a}$H.Niewodnicza\'nski Institute of Nuclear Physics, Polish Academy of
Sciences,\\
ul. Eliasza-Radzikowskiego 152, PL 31342 Krak\'ow, Poland\\
e-mail: pawel.blasiak@ifj.edu.pl, agnieszka.gawron@ifj.edu.pl,
andrzej.horzela@ifj.edu.pl}
\authorii{K. A. Penson$^{b}$ and A. I Solomon$^{b,c}$}
\addressii{$^{b}$Laboratoire de Physique Th\'eorique de la Mati\`ere
Condens\'ee,
Universit\'e P. 
et M. Curie,\\ Tour 24 - 2e \'et., 4 Pl.Jussieu, F 75252 Paris Cedex 05, France\\
$^{c}$Physics and Astronomy Department, The Open University,\\
Milton Keynes MK7 6AA, United Kingdom
\\email: penson@lptl.jussieu.fr, a.i.solomon@open.ac.uk}
\authoriii{}      \addressiii{}
%
\headauthor{P. B\l{}asiak et al.}
\headtitle{Combinatorial Solutions to Normal Ordering of Bosons}
\lastevenhead{P. B\l{}asiak et al.: Combinatorial Solutions to Normal Ordering of Bosons}
\maketitle
\begin{abstract}
We present a combinatorial method of constructing solutions to the normal ordering of boson operators. Generalizations of standard combinatorial notions - the Stirling and Bell numbers, Bell polynomials and Dobi\'nski relations - lead to calculational tools which allow to find explicitly normally ordered forms for a large class of operator functions. 
\end{abstract}
\section{Preface}
Solutions to mathematical problems which we would call nowadays examples of
solutions to the normal ordering problem have been known from a long time. 
Transformation of a differential operator $(x\frac{\rm d}{{\rm d}x})^n$ into a form
in which all derivatives stand to the right with respect to all powers of
$x$, or the Baker-Campbell-Hausdorff formula which allows to disentangle the
exponential operator $\exp{(A+B)}$ for operators $A$ and $B$ satisfying
commutation relations $[A,B]=C, [A,C]=0, [B,C]=0$ provide us with well known examples. In the framework of the second quantization scheme normal ordering of expressions involving ladder operators, {\it i.e.} transformation of such expressions into a form in which all creation operators are moved to the left with respect to all annihilation operators using appropriate commutation relations, plays an important role as it simplifies evaluation of the vacuum and coherent state matrix elements in quantum field theory, many body quantum physics and quantum optics. A standard way of reaching the goal is provided by the Wick theorem - an universal tool, unfortunately ineffective when one wants to investigate general structure of solutions for a vast majority of physical problems. Thus it is highly desirable to find manageable formulas or at least some guiding principles leading to the solutions of general normal ordering problem formulated as follows: consider an operator $F(a^{\dagger},a)$ depending on canonical boson ladder operators $[a,a^{\dagger}]=1$ and search for solutions $G(a^{\dagger},a)$ of a functional operator equation
\begin{equation}
\label{general}
F(a^{\dagger},a) \equiv N[F(a^{\dagger},a)] = :G(a^{\dagger},a):
\end{equation}
\noindent where normally ordered operator $N[F(a^{\dagger},a)]$ is obtained using nontrivial commutation relations while under the symbol $:\!~\!:$ the ladder operators are considered as commuting ones. Research on  general normal ordering problem started from seminal paper of Cahill and Glauber \cite{cahill}  who considered operators bounded with respect to the Hilbert-Schmidt norm, expanded them in terms of ordered products of ladder operators and found integral representations for coefficients of such expansions. The method, however, allowed to solve (\ref{general}) explicitly only for the simplest examples. Investigations were pushed forward when Navon \cite{navon} and Katriel \cite{katriel} showed that the normal ordering problem is situated firmly in the area of combinatorics and that methods of combinatorial analysis do appear effective in finding solutions of (\ref{general}). Since that time many examples have been studied and normally ordered forms of operators depending on $(a^{\dagger})^ra^s$ characterized by integer powers $r$ and $s$ or depending on $q(a^{\dagger})a + v(a^{\dagger})$ with arbitrary functions $q$ and $v$ were found \cite{blasiak1} - \cite{blasiak4}. Here we are going to present basic concepts of our approach and to sketch combinatorial methods useful in order to solve the boson normal ordering problem. 
\section{Stirling and Bell numbers, Dobi\'nski relations - the generic example}
Considering canonical boson ladder operators $a$ and $a^{\dagger}$ we can ask
what is normally ordered expression for the $n$-th power of the number operator $N=a^{\dagger}a$.
Obviously such an expression may be written down as
\begin{equation}
\label{stirlings1}
N^n = (a^{\dagger}a)^n = \sum_{k=1}^n S(n,k) (a^\dag)^k a^k
\end{equation}
\noindent  but in order to solve the problem one has to determine 
$S(n,k)$'s. This is easy to do using mathematical induction - the numbers $S(n,k)$ satisfy the recurrence
\begin{equation}
\label{stirlings2}
S(n+1,k)=kS(n,k)+S(n,k-1)
\end{equation}
\noindent 
which defines the \emph{Stirling numbers of the second kind} - positive integers which appear in enumerative combinatorics \footnote{The Stirling numbers of the second kind count the number of ways of putting $n$ different objects into $k$ identical containers none leaving empty} or in classical problems:
\begin{equation}
\label{stirlings3}
\begin{array}{ccc}
\left(x\frac{\displaystyle \rm d}{\displaystyle {\rm d}x}\right)^{n} 
= \sum\limits_{k=1}^{n}S(n,k)x^k 
\frac{\displaystyle \rm d^{k}}{\displaystyle {\rm d}x^{k}},
&~~~~~&
x^n=\sum\limits_{k=1}^n S(n,k) x^{\underline{k}},
\end{array}
\end{equation}
\noindent  where  $x^{\underline {k}}=x\cdot(x-1)\cdot...\cdot(x-k+1)$ denotes the
\emph{falling factorial} ($x^{\underline {0}}=1$). The second equation in  (\ref{stirlings3}), called the \emph{Stirling transform}, describes transformation of basis in the space of polynomials so the Stirling numbers of the second kind may be also interpreted as coefficients leading from $\{x^{\underline{n}}\}_{n=0}^\infty$ to
$\{x^n\}_{n=0}^\infty$, while the \emph{Stirling numbers of the first kind} provide us with coefficients of the reverse transformation.

\noindent The basic tool extensively used in our further considerations is 
the Dobi\'nski formula. In order to derive it let us apply (\ref{stirlings3})
to the exponential $e^x$.  This gives
\begin{equation}
\label{stirlings6}
e^{-x}\sum_{k=0}^\infty k^n\frac{x^k}{k!} = \sum_{k=1}^n S(n,k) x^k = B(n,x),
\end{equation}
\noindent namely the Dobi\'nski formula, which is a remarkable relation 
between some polynomials  
$B(n,x)$ (called Bell, or exponential, polynomials \cite{comtet}) and products of infinite series.
List of implications of (\ref{stirlings6}) reads

\noindent --- positive integers $B_n = B(n,1)$, called Bell numbers, can be written for any $n$ as infinite sums of fractions

\begin{equation}
\label{stirlings7}
B(n)=\frac{1}{e}\sum_{k=1}^\infty \frac{k^n}{k!},
\end{equation}

\noindent --- the Stirling numbers of the second kind can be explicitly calculated from

\begin{equation}
\label{stirlings8}
S(n,k)=\frac{1}{k!}\sum\limits_{j=1}^k\left(\begin{array}{c}{k}\\{j}\end{array}\right)(-1)^{k-j}j^n,
\end{equation}

\noindent --- a closed formula for exponential generating function of the Bell polynomials can be obtained

\begin{equation}
\label{stirlings9}
\begin{array}{rcl}
G(\lambda,x)
&=&
\sum\limits_{n=0}^\infty B(n,x)\frac{\lambda^n}{n!}
=
e^{-x}\sum\limits_{n=0}^\infty \sum\limits_{k=0}^\infty
k^n\frac{\displaystyle x^k}{\displaystyle k!}\frac{\displaystyle \lambda^n}{\displaystyle n!}
\\
&=&e^{-x} \sum\limits_{k=0}^\infty
\frac{\displaystyle x^k}{\displaystyle k!}\sum\limits_{n=0}^\infty k^n\frac{\displaystyle \lambda^n}{\displaystyle n!}
=e^{-x} \sum\limits_{k=0}^\infty \frac{\displaystyle x^k}{\displaystyle k!}e^{\lambda k}
= e^{x(e^\lambda-1)}.
\end{array}
\end{equation}

\noindent Using the properties of standard coherent states (defined as eigenstates of the annihilation operator, $a|z\rangle=z|z\rangle$) we conclude that the diagonal coherent state
matrix elements of $N^n$ generate Bell polynomials:

\begin{equation}
\label{stirlings10}
\begin{array}{c}
\langle z|(a^\dag a)^n|z\rangle =\langle z|\sum\limits_{k=1}^n S(n,k){a^{\dag}}^k a^k|z\rangle 
=\sum\limits_{k=1}^n S(n,k) |z|^{2k} = B(n,|z|^2).
\end{array}
\end{equation}

\noindent Moreover, expanding the exponential $e^{\lambda a^\dag a}$ (with $\lambda$ being an arbitrary parameter) and
taking the diagonal coherent state matrix element we have

\begin{equation}
\label{stirlings11}
\begin{array}{c}
\langle z|e^{\lambda a^\dag a}|z\rangle
=\sum\limits_{n=0}^\infty\langle z|(a^\dag
a)^n|z\rangle\frac{\displaystyle \lambda^n}{\displaystyle n!}
=\sum\limits_{n=0}^\infty
B(n,|z|^2)\frac{\displaystyle\lambda^n}{\displaystyle n!},
\end{array}
\end{equation}

\noindent which means that the diagonal coherent state matrix element of
$e^{\lambda a^\dag a}$ yields the exponential generating function
of the Bell polynomials
$\langle z|e^{\lambda a^\dag a}|z\rangle=e^{|z|^2(e^\lambda-1)}$.
Relation between the considerations above and the normal ordering problems comes from
an important property of the coherent state representation \cite{louisell}: 

{\it If for an  arbitrary operator ${F}(a,a^{\dag})$ we have
\begin{equation}
\label{stirlings13}
\begin{array}{l}
\langle z'|{F}(a,a^{\dag})|z\rangle = \langle z^\prime|z\rangle\ G(z^{\prime *},z)
\end{array}
\end{equation}

then the normally ordered form of ${F}(a,a^{\dag})$ is given by}

\begin{equation}
\label{stirlings14}
\begin{array}{l}
{\cal N}\left[{F}(a,a^{\dag})\right] =\, :G(a^{\dag},a):~.
\end{array}
\end{equation}

\noindent Consequently, the knowledge of the generating function of the Bell polynomials lead to 

\begin{equation}
\label{stirlings15}
e^{\lambda a^\dag a}={\cal N}\left[e^{\lambda a^\dag
a}\right]\equiv\ :e^{a^\dag a(e^\lambda-1)}\ :,
\end{equation}

\noindent {\it i.e.} the normally ordered form of
$e^{\lambda a^\dag a}$. We shall generalize this result, using the methods just described, to a quite large class of operator functions. 

\section{Normal Ordering of Homogeneous Boson Polynomials}

Let us treat the normal ordering of expressions in boson
ladder operators called homogeneous boson polynomials and given as linear combinations of monomials
\begin{equation}
\label{nobe1}
(a^\dag)^{r_M}a^{s_M}\dots(a^\dag)^{r_2}a^{s_2}(a^\dag)^{r_1}a^{s_1}, 
\end{equation}
\noindent all having the same difference between the total
number of creation and annihilation operators which we call excess, denoted by $d=\sum_{i=1}^{M}(r_{i}-s_{i})$ and assumed to be nonnegative. 
It means that we consider operators which normal form is
\begin{equation}
\label{nobe2}
H_{\boldmath{\alpha}}^d=(a^\dag)^d\ \sum_{k=N_0}^N\alpha_k\
(a^\dag)^k a^k,
\end{equation}
\noindent with $N_0=s_1$, $N=\sum_{i=1}^{M}s_i$ and $\alpha_k$'s to be specified for any initially given linear combination of monomials (\ref{nobe1}) \cite{blasiak5},\cite{blasiak6}.

\noindent Calculating the normally ordered form of
$(H_{\boldmath{\alpha}}^d)^n$ we get
\begin{equation}
\label{nobe3}
\left(H_{\boldmath{\alpha}}^d\right)^n=(a^\dag)^{nd}\ \sum_{k=N_0}^{nN}S_{\boldmath{\alpha}}^d(n,k)\ (a^\dag)^k a^k,
\end{equation}
\noindent with $S_{\boldmath{\alpha}}(n,k)$ to be determined. These
coefficients obviously generalize the Stirling numbers, provide us with generalized 
Bell polynomials and Bell numbers
\begin{equation}
\label{nobe4}
\begin {array}{rcl}
B_{\boldmath{\alpha}}^d(n,x)=\sum\limits_{k=N_0}^{nN}S_{\boldmath{\alpha}}^d(n,k)\
x^k,
&~~&
B_{\boldmath{\alpha}}^d(n)=B_{\boldmath{\alpha}}^d(n,1)=\sum\limits_{k=N_0}^{nN}S_{\boldmath{\alpha}}^d(n,k)
\end{array}
\end{equation}
\noindent and share general properties of the standard Stirling and Bell numbers \cite{blasiak5},\cite{blasiak6}:

\noindent --- {satisfy the recurrence relation generalizing (\ref{stirlings2})}
\begin{equation}
\label{nobe6}
S_{\boldmath{\alpha}}^d(n+1,k)=\sum_{l=N_0}^N\alpha_l\
\sum_{p=0}^l \left(\begin{array}{c}{l}\\{p}\end{array}\right)(nd+k-l+p)^{\underline{p}}\ S_{\boldmath{\alpha}}^d(n,k-l+p),
\end{equation}
--- {determine connection between two sets of polynomials}
\begin{equation}
\label{nobe7}
\prod_{i=1}^n\sum\limits_{k=N_0}^N\alpha_k\ (x+(i-1)d)^{\underline{k}}=\ \sum\limits_{k=N_0}^{nN}S_{{\boldmath\alpha}}^d(n,k)\ x^{\underline{k}},
\end{equation}
--- {lead to generalized Dobi\'nski-type relations}
\begin{equation}
\label{nobe8}
B_{{\boldmath\alpha}}^d(n,x)=e^{-x}\sum\limits_{l=0}^\infty\left[\prod_{i=1}^n
\sum\limits_{k=N_0}^N\alpha_k\
(l+(i-1)d)^{\underline{k}}\right]\frac{x^l}{l!}.
\end{equation}
\noindent Generalized Stirling numbers and their generating functions are explicitly given by
\begin{equation}
\label{nobe9}
\begin{array}{c}
S_{{\boldmath\alpha}}^d(n,k)=\frac{1}{k!}
\sum\limits_{j=0}^k\left(\begin{array}{c}{k}\\{j}\end{array}\right)(-1)^{k-j}
\prod\limits_{i=1}^n\sum\limits_{l=N_0}^N\alpha_l\
(j+(i-1)d)^{\underline{l}},\\
\sum\limits_{n=k}^\infty\!S_{\boldmath{\alpha}}^d(n,k) \frac{\lambda^n}{n!}
=\frac{1}{k!}\sum\limits_{l=0}^k\!\left(\begin{array}{c}{k}\\{l}\end{array}\right)\! 
(-1)^{k-l}
\sum\limits_{n=0}^\infty\left[\,\prod\limits_{i=1}^n\sum\limits_{k=N_0}^N\!\alpha_k (l+(i-1)d)^{\underline{k}}\!\right]\frac{\lambda^n}{n!}
\end{array}
\end{equation}
\noindent from which it is seen that generalized Stirling numbers can be expressed in terms of ordinary Stirling numbers $S(m,l)$ with $m = 1,\cdots, n(N-N_{0}+1)$.

\noindent If we introduce exponential generating functions of generalized Bell
polynomials
\begin{equation}
\label{nobe10}
G_{\boldmath{\alpha}}^d(\lambda,x)=\sum_{n=0}^\infty
B_{\boldmath{\alpha}}^d(n,x)\frac{\lambda^n}{n!}
\end{equation}
\noindent then we can derive relations between coherent state matrix elements and suitable generating functions. Analogously to (\ref{stirlings11}) we have
\begin{equation}
\label{nobe11}
\langle z|e^{\lambda H_{\boldmath{\alpha}}^d}|z\rangle
=\sum_{n=0}^\infty\langle
z|\left(H_{\boldmath{\alpha}}^d\right)^n|z\rangle\frac{\lambda^n}{n!}
=\sum_{n=0}^\infty \
B_{\boldmath{\alpha}}^d(n,|z|^2)\frac{\left((z^*)^d\lambda\right)^n}{n!}
\end{equation}
\noindent which yields
$\langle z|e^{\lambda H_{\boldmath{\alpha}}^d}|z\rangle
=G_{\boldmath{\alpha}}^d\left((z^*)^d\lambda,|z|^2\right)$
and 
\begin{equation}
\label{nobe13}
e^{\lambda H_{\boldmath{\alpha}}^d} =\
:G_{\boldmath{\alpha}}^d((a^\dag)^d\lambda,a^\dag a):\ .
\end{equation}
\noindent Inserting the generalized Dobi\'nski relation (\ref{nobe8}) into the generating function (\ref{nobe10}) one obtains the latter as
\begin{equation}
\label{nobe14}
G_{\boldmath{\alpha}}^d(\lambda,x) =e^{-x}\sum_{n=0}^\infty
\sum_{l=0}^\infty\left[\prod_{i=1}^n\sum_{k=N_0}^N\alpha_k\
(l+(i-1)d)^{\underline{k}}\right]\frac{x^l}{l!}\frac{\lambda^n}{n!}
\end{equation}
\noindent and,  after a change of summation order, as
\begin{equation}
\label{nobe15}
G_{\boldmath{\alpha}}^d(\lambda,x)=e^{-x}\sum_{l=0}^\infty\frac{x^l}{l!}\sum_{n=0}^\infty
\left[\prod_{i=1}^n\sum_{k=N_0}^N\alpha_k\
(l+(i-1)d)^{\underline{k}}\right]\frac{\lambda^n}{n!}.
\end{equation}
\noindent The series (\ref{nobe15}) is convergent either 
if $N_{0}=N=1$ which means that we are treating operators including arbitrary number of creators and only one annihilator or if $d=0$, which corresponds to $H^{0}_{\alpha}$ depending on the number operator. The first case may be illustrated on the example of $(a^{\dagger})^{r}a$,  \cite{blasiak2}, and generalized to the normal ordering problems of operators involving expressions being linear in $a$ and depending on arbitrary functions $q(a^{\dagger})$ and $v(a^{\dagger})$
\begin{equation}
\label{nobe16}
\begin{array}{c}
{\cal N}\left[e^{\lambda\left[q(a^{\dag})a +
v(a^{\dag})\right]}\right] =
:g(\lambda,a^{\dag})e^{\left[T(\lambda,a^{\dag}) -
a^{\dag}\right]a}:\ ,
\end{array}
\end{equation}
\noindent where, according to the substitution theorem, both functions $g$, $T$ are unique solutions of linear differential equations determined by $q$ and $v$.
Coherent state matrix elements of (\ref{nobe16}) give generating functions of Sheffer polynomials \cite{blasiak3}. For the second case the internal sum becomes exponential of a polynomial in  $l$ and the external sum converges if $\alpha_N\lambda < 0$. This leads, \cite{blasiak4}, to
\begin{equation}
\label{nobe17}
G_{\boldmath{\alpha}}^d(\lambda,x) =e^{-x}\sum_{l=0}^\infty
\exp{\left(\lambda\sum_{k=N_0}^N\alpha_k\
l^{\underline{k}}\right)}\frac{x^l}{l!}.
\end{equation}

\noindent  Combinatorial generating functions (\ref{nobe10}), introduced as formal power series, may be divergent but if we require their interpretation as physically meaningful matrix elements then we have to give them analytical meaning, at least in the sense of asymptotic expansions. It is our conjecture that in such a case the Pad\'e method of generalized summation may be successfully applied to series (\ref{nobe10}). If we know well-defined solutions (like (\ref{nobe17}) is) then Pad\'e summation results may be compared with them - it appears that the agreement is unexpectedly good and we expect that this occurs for another cases as well.   

{\small A.G. wishes to thank the Polish Ministry of Scientific Research and
Information Technology for support under Grant No: 1P03B 060 27.}

\end{document}